\documentclass[a4paper,11pt]{article}

\pdfoutput=1
\usepackage{jcappub}

\usepackage{graphicx}
\usepackage{longtable}
\usepackage{float}
\usepackage{dcolumn}
\usepackage{bm}
\usepackage{appendix}
\usepackage{multirow}
\usepackage{color}
\usepackage[utf8]{inputenc}
\usepackage{footmisc}
\usepackage{hyperref}

\usepackage{xcolor}
\usepackage{graphicx}
\usepackage{bm}
\usepackage{ulem}
\usepackage{multirow, array, float, tabularx}
\usepackage{amsmath}
\hypersetup{
	citecolor = red,
	linkcolor = blue,
	urlcolor = magenta
}

\newcommand{\be}{\begin{equation}}
	\newcommand{\ee}{\end{equation}}
\newcommand{\bea}{\begin{eqnarray}}
	\newcommand{\eea}{\end{eqnarray}}
\newcommand{\bes}{\begin{subequations}}
	\newcommand{\ees}{\end{subequations}}

\newcommand{\bc}{\begin{center}}
	\newcommand{\ec}{\end{center}}

\usepackage{wasysym}

\begin{document}

\title{Forecasting constraints on quintessential inflation from future\\ generation of galaxy and CMB surveys}

\author[a]{G. Rodrigues}\emailAdd{gabrielrodrigues@on.br}
\author[a]{F. B. M. dos Santos}\emailAdd{fbmsantos@on.br}
\author[b,c]{S. Santos da Costa}\emailAdd{simony.santosdacosta@unitn.it}
\author[a]{J. G. Rodrigues}\emailAdd{jamersonrodrigues@on.br}
\author[d]{R. von Marttens}\emailAdd{rodrigomarttens@ufba.br}
\author[e,f,g]{R. Silva}\emailAdd{raimundosilva@fisica.ufrn.br}
\author[h]{D. F. Mota}\emailAdd{d.f.mota@astro.uio.no}
\author[a]{J. S. Alcaniz}\emailAdd{alcaniz@on.br}

\affiliation[a]{Observatório Nacional, Rio de Janeiro - RJ, 20921-400, Brasil}
\affiliation[b]{Department of Physics, University of Trento, Via Sommarive 14, 38123 Povo (TN), Italy}
\affiliation[c]{Trento Institute for Fundamental Physics and Applications (TIFPA)-INFN, Via Sommarive 14, 38123 Povo (TN), Italy}
\affiliation[d]{Instituto de Física, Universidade Federal da Bahia, Salvador - BA, 40210-340, Brasil}
\affiliation[e]{Departamento de Física, Universidade Federal do Rio Grande do Norte, Natal - RN, 59072-970, Brasil}
\affiliation[f]{National Institute of Science and Technology of Complex Systems, Brazilian Center for Physics Research, Rio de Janeiro - RJ, 22290-180, Brazil}
\affiliation[g]{Departamento de F\'{\i}sica, Universidade do Estado do Rio Grande do Norte, Mossor\'o - RN, 59610-210, Brasil}
\affiliation[h]{Institute of Theoretical Astrophysics, University of Oslo, Oslo, P.O.Box 1029 Blindern, N-0315, Norway}

\abstract{
We investigate the constraining power of future CMB and galaxy surveys on models of quintessential inflation realized within the framework of $\alpha$-attractors. We analyze how these future datasets will probe the parameter space of $\alpha$-attractor quintessential inflation, specifically the inflationary potential parameters. Our results demonstrate that the synergy between CMB-S4, LiteBIRD, and Euclid can significantly tighten the bounds on the model parameters, achieving forecasted $1\sigma$ uncertainties of $\alpha=2\pm 0.17$,  $n_s=0.965\pm 0.0014$, $\ln(10^{10}A_s)=3.0447\pm 0.0029$ for the CMB+GC$_{sp}$ case. This level of sensitivity will enable us to discriminate between different realizations of quintessential inflation and test the attractor behavior characteristic of these models.}

\maketitle

\section{Introduction}

In the standard cosmological framework, it is generally accepted that the universe has undergone two periods of accelerated expansion: one in very early times and the other in the present era. These two phases, known as early and late-time cosmic acceleration, are understood as important conditions for a consistent description of cosmic history~\cite{Planck:2018nkj,Planck:2018vyg}. On the other hand, the causes that lead to both descriptions are generally very different. In early times, the universe was dominated by a scalar field, whose potential allows a regime in which the energy density varies slowly enough to be almost constant, resulting in an accelerated expansion \cite{Baumann:2009ds}. This is the basis of the inflationary scenarios~\cite{Starobinsky:1980te,Guth:1980zm,Linde:1981mu}, where this field would have its remaining energy dissipated into the standard model particles through a reheating mechanism, thus realizing the transition to the radiation-dominated era \cite{Abbott:1982hn,Albrecht:1982mp,Kofman:1994rk,Kofman:1997yn}.

In late times, the universe is well described by the $\Lambda$CDM model, in which the component behind cosmic acceleration is the cosmological constant \cite{Weinberg:1988cp,Peebles:2002gy,Padmanabhan:2002ji}. There
is no consensus on the fundamental nature of this dark
energy component, or even its fraction that dominates the present energy budget. Moreover, the issue of cosmic tensions, such as the one involving the present value of the Hubble parameter between early and late-time probes~\cite{Freedman:2017yms,Verde:2019ivm,DiValentino:2021izs}, could indicate new physics beyond the standard model.

A possible option that can be considered to properly describe both early and late-time epochs, is the use of an unified approach, in which one can assume that both periods of acceleration are caused by the same agent, a scalar field that initially obeys the conditions to realize early inflation, which afterwards, has its energy density diluted enough to behave as dark energy in the present \cite{Peebles:1998qn,Peloso:1999dm,Dimopoulos:2001ix,Tashiro:2003qp}. This unified picture is dubbed \textit{quintessential inflation} (QI), whose viability has been investigated with cosmological data recently \cite{Geng:2015fla,Dimopoulos:2017zvq,Dimopoulos:2017tud,Geng:2017mic,Akrami:2017cir}. Naturally, such a scenario imposes a constraint on the shape of the potential function $V(\phi)$, which should be flat enough at early times, to ensure the inflationary period, while having, for instance, a \textit{runaway} behavior that could allow the chosen model to behave as a quintessence one.

In this work, we investigate the prospects of the quintessential inflation scenario, considering configurations from future large-scale structure (LSS) and cosmic microwave background (CMB) surveys. In particular, we focus on the Euclid survey~\cite{euclid24}, combined with the next-generation CMB experiments CMB-S4 and LiteBIRD.
This extends the work done in \cite{Akrami:2020zxw}, in which the constraining power of future LSS surveys was projected through a Fisher matrix analysis. For both galaxy clustering and weak lensing analyses, due to the nature of the model, a significant reduction in the uncertainty of the scalar spectral index $n_s$ is achieved when compared to projections for the $\Lambda$CDM model. This also indicates that the combination of both CMB and LSS data can provide even more significant constraints on the relevant parameters.

This motivates the present discussion. Following the work of \cite{Akrami:2017cir,Akrami:2020zxw}, we choose the potential they studied, which is motivated by supergravity, in the context of $\alpha$-attractor inflation. A great advantage of these classes of models is that in the inflationary description, the predictions on the cosmological quantities are dictated by the choice of the parameter $\alpha$, having the universal feature of driving the spectral index $n_s$ and the tensor-to-scalar ratio $r$ to specific values, in the limit of low $\alpha$. We note that recently, numerical analyses with current cosmological data of quintessential inflation were carried out in \cite{Giare:2024sdl,Alestas:2024eic}, for the model we investigate here. In this way, we seek the possibility of future CMB and LSS data to establish the capability of constraining the central parameter $\alpha$, which will help us characterize this specific scenario, as the $\alpha$ values are correlated with the underlying theoretical background that motivates the $\alpha$-attractor inflation.

This work is organized in the following manner: In Section \ref{sec2}, we give a general review of quintessential inflation models and which are the requirements for their viability; we also introduce the specific model to be investigated. In Section \ref{sec3}, we present the methodology used to perform the forecast analysis, as well as the features of each of the future surveys considered. We discuss the results in Section \ref{sec4}, while also presenting our conclusions.

\section{The $\alpha$-attractor quintessential inflation}\label{sec2}

We start from the picture of a cosmological scalar field, whose Lagrangian density is given by
\begin{equation}
    \mathcal{L}=\sqrt{-g}\left(\frac{M_P^2}{2}R - \frac{1}{2}\partial_\mu\phi\partial^\mu\phi - V(\phi)\right),
    \label{eq:1}
\end{equation}
with $M_P=1/\sqrt{8\pi G}$ being the reduced Planck mass, while $R$ is the Ricci scalar. The dynamics of the inflaton field $\phi$ is dictated by the choice of potential $V(\phi)$, which must be flat enough such that the potential energy changes slowly with time, as the kinetic energy of the field is subdominant. This is the so-called \textit{slow-roll} regime, in which the slow-roll parameters
\begin{equation}
    \epsilon\equiv\frac{M_P^2}{2}\left(\frac{V_{,\phi}}{V}\right)^2, \quad \eta\equiv M_P^2\frac{V_{,\phi\phi}}{V},
\end{equation}
must be smaller than one during inflation, such that the accelerated regime can take place. Here we used the notation $V_{,\phi}\equiv dV(\phi)/d\phi$ and  $V_{,\phi\phi}\equiv d^2V(\phi)/d\phi^2$.

The main observables related to the inflationary period are the spectral index $n_s$ and the tensor-to-scalar ratio $r$, respectively given by
\begin{equation}
    n_s = 1 - 6\epsilon + 2\eta, \quad r=16\epsilon.
\end{equation}
CMB experiments measure both quantities at a certain pivot scale $k_\star$, at which the relevant scales for inflation leave the horizon during the accelerated expansion. In particular, the spectral index comes from the variation of the primordial power spectrum $P(k)$ as a function of scale, such that the amplitude of the scalar perturbations at the pivot scale can be related to the scalar field quantities as
\begin{equation}
    P(k_\star) \equiv A_s=\frac{V_\star}{24\pi^2M_P^4\epsilon_\star},\label{eq:Amp_S}
\end{equation}
which is well determined by Planck as $\ln(10^{10} A_s)=3.044\pm 0.014$ \cite{Planck:2018nkj}. 

One important class of inflationary models is the one given by $\alpha$-attractors. They are usually characterized by the non-canonical Lagrangian density,
\begin{equation}
    \mathcal{L}=\sqrt{-g}\left(\frac{M_P^2}{2}R - \frac{1}{2}\frac{\partial_\mu\phi\partial^\mu\phi}{\left(1-\frac{\phi^2}{6\alpha}\right)^2} - V(\phi)\right).
\end{equation}
This kind of model arises from the supersymmetric constructions of the theory of gravity, dubbed collectively as $\alpha-$attractor theories \cite{Kallosh:2013hoa,Kallosh:2013yoa,Kallosh:2014rga}. In common, all these realizations of supergravity present a non-canonical kinetic term for the scalar field $\phi$, with a characteristic pole at $\phi^2 = 6\alpha$, defined by the curvature of the field manifold in the K\"{a}hler potential \cite{Kallosh:2013yoa}. Such a pole structure perfectly fits the conditions for QI potential, enabling flat directions at large amplitudes of the canonical field, $|\varphi| \rightarrow \infty$, controlled by the magnitude of $\alpha$. One convenient feature of $\alpha$-attractor models is that one can work in a setting where the field can be canonically normalized. This is done by a redefinition given by
\begin{equation}\label{eq:7}
    \phi\equiv\sqrt{6\alpha}\tanh\frac{\varphi}{\sqrt{6\alpha}},
\end{equation}
which drives the poles to $\varphi\rightarrow\pm\infty$. This is another nice feature present in this construction, as it allows flat directions in the field even for an arbitrary potential \cite{Linde:2016uec}, thus allowing the global behavior in the predictions of the inflationary parameters, being
\begin{equation}
    n_s\simeq 1-\frac{2}{N_\star}, \quad r\simeq\frac{12\alpha}{N_\star^2},
    \label{eq:8}
\end{equation}
where $N_\star$ is the number of e-folds, $N \equiv\ln(a)$, from the horizon crossing up to the end of inflation. 

The discussion presented in \cite{Akrami:2017cir} shows that even simple potentials such as the linear and exponential ones, usually not suited for inflation by either their own construction, or the exclusion by CMB data (see however \cite{Rodrigues:2020fle} for a working example with the quartic potential $V(\phi)=\lambda\phi^4/4$), can become viable and a reasonable description of early-time physics. The most interesting feature, however, arises when the shape of $V(\varphi)$ allows an inflationary plateau on one side, while the other presents a quintessential tail. This allows the construction of quintessential inflation models, as an attempt to unify the description of inflation and the current state of acceleration of the universe by the action of a single scalar field.

Here, we consider one of the simplest forms for the potential, being the exponential one 
\begin{equation}
    V(\phi) = M^2 e^{\gamma\left(\phi/\sqrt{6\alpha}-1\right)}.
\end{equation}
The model has the following parameters: $M^2$, in units of $M_P^4$, and the dimensionless parameter $\gamma$. As discussed in \cite{AresteSalo:2021wgb,Dimopoulos:2017tud}, by setting eq. \ref{eq:7} into the potential, we will find
\begin{equation}
    V(\varphi) = M^2 e^{\gamma\left(\tanh\varphi/\sqrt{6\alpha}-1\right)}.
    \label{eq:13}
\end{equation}

Inflation proceeds on the positive field side of the potential, as $+\varphi\gg1$ leads to $V(|\varphi|\gg 1)^+ \rightarrow M^2$. Eventually, the field will roll down the potential to negative values, such that in the large field limit, we find $V(|\varphi|\gg 1)^-\rightarrow M^2 e^{-2\gamma}$. We then note that in the future, the potential can behave nearly as a cosmological constant, given the right conditions. Here we come back to the value of the parameter $M^2$: while in regular quintessence models, $M^2$ is usually entirely constrained by late-time data, here this parameter is linked to the early period. Thus, we can expect that additional constraints coming from the energy scale difference between inflation and the present will determine the viability of this or any quintessential inflation model. Furthermore, given the nature of the potential and the balance between energy scales at early and late times, models of this type are usually referenced as providing a \textit{cosmological seesaw} mechanism, in which the unified picture can explain the difference in scales in a natural manner.

The amplitude of scalar fluctuations $A_s$ is connected with the inflationary potential through \ref{eq:Amp_S}, in a way that one of the free parameters of the model can be tightly constrained, given the restrictions on $A_s$ itself. In this case, the energy scale of the potential can be related to $A_s$ as \cite{Akrami:2017cir}
\begin{equation}
    \frac{M^2}{M_P^4}=\frac{144\pi^2\alpha N_\star A_s}{(2N_\star - 3\alpha)^3},
\end{equation}
setting the energy density of the inflationary process to the order of $\sim 10^{-10}M_P^4$, once considering the measurements of the associated observables. As usual, the early inflationary dynamics can be seen as the result of a high-energy process. The late-time expansion, on the other hand, is a low-energy process, with observations pointing to $\rho_\Lambda \sim 2.5 \times 10^{-11}$ eV$^4$ \cite{Planck:2018vyg}. To align the predictions of the early- and late-time energy densities, one should fit another parameter, in our case $\gamma$, to obtain the correct steepness for the proposed potential, ensuring a consistent transition from the inflationary period to the late-time expansion.

Before moving forward, we briefly discuss which results from the observational point of view we have so far. In \cite{Akrami:2017cir}, a background analysis was carried out to establish the viable parameter space of the model. In their Fig. 15, there is a clear correlation between $\gamma$ and $M^2$, highlighting how early and late-time quantities must relate to having a consistent cosmic history. In a later study \cite{Akrami:2020zxw}, predictions from future surveys were conducted, exploring how we could differentiate these models from the standard cosmological scenario. This happens thanks to the dependence of the model on primordial parameters, since, as we can see in their Fig. 3, the difference in the Hubble expansion rate for different values of $\alpha$ is small, indicating a need for higher precision of future cosmological data. More recently, in \cite{Alestas:2024eic}, the model is properly solved in the full scalar field equations, while also considering the inflationary effect through the $n_s$ and $r$ parameters, as in Eq. \ref{eq:8}.

\section{Methodology and mock likelihoods}\label{sec3}

We perform a Monte Carlo Markov Chain (MCMC) analysis of simulated data to forecast the sensitivity of future CMB and LSS experiments to the parameters of the quintessential inflation scenario described in Sec.~\ref{sec2}. Although the simplicity of the Fisher matrix \cite{Vogeley:1996xu,Tegmark:1996bz,coe:2009fis} formalism is a notable advantage of this method, the possible non-Gaussian dependence of the likelihoods on the parameters of the cosmological model may jeopardize the reliability of its results~\cite{Perotto:2006rj}, and justifies our choice in this work.

The $\mathrm{MontePython}$ \cite{Audren:2012wb,Brinckmann:2018cvx} code is used to sample the likelihood of the mock data, following the Metropolis-Hastings algorithm. The specifications chosen for each of the likelihoods are detailed below. To achieve a faster convergence of the chains, the covariance matrix obtained with the Fisher approach was introduced as input for the proposal function. The covariance matrix is then updated until the Gelman-Rubin convergence criterion achieves a value of $R - 1 < 0.01$ for every parameter. In parallel, the Cosmic Linear Anisotropy Solving System ($\mathrm{CLASS}$) \cite{Diego_Blas_2011} was modified in order to solve the Boltzmann equations for the $\alpha-$attractor quintessential model and compute the cosmological observables. 

The diversity of cosmological probes at different cosmic epochs and distance scales allows one to give prospects on the future capabilities of many cosmological surveys, consequently providing the best restrictions on distinct models. Specifically, the combination of results from future galaxy and CMB surveys might be the key to determining cosmological parameters and the possibility of physics beyond the standard model. In particular, when combining galaxy clustering data, which can provide high-precision measurements of $n_s$~\cite{Akrami:2020zxw}, and therefore, information regarding the nature of inflation and its dynamics, with CMB experiments measuring $r$ (the amount of primordial gravitational waves), through detections of its B polarization mode, can potentially be able to determine the value of $\alpha$, providing a test of $\alpha$-attractors models against their non-quintessential versions.

\subsection{CMB Stage-IV}

Many CMB surveys are expected to be operational in the next decade, such as the CMB-S4 collaboration \cite{CMB-S4:2016ple,CMB-S4:2020lpa}, the LiteBIRD satellite \cite{Matsumura:2013aja,Hazumi:2019lys,LiteBIRD:2022cnt}. We consider a CMB-S4 setup plus a LiteBIRD contribution for the lower multipoles of the CMB spectra. In essence, we consider the impact of the temperature (T) and polarization modes E and B, which can be related through the $\textbf{C}_\ell$ matrix
\begin{equation}
   \textbf C_\ell = \begin{pmatrix}
C_\ell^{TT} + N_\ell^{TT} & C_\ell^{TE} & 0\\
C_\ell^{TE} & C_\ell^{EE} + N_\ell^{EE} & 0\\
0 & 0 & C_\ell^{BB} + N_\ell^{BB}\\ 
\end{pmatrix}\;,
\end{equation}
in which $C_\ell$ corresponds to the coefficients of the temperature, polarization, and cross-correlation power spectra, while $N_\ell$ represents the instrumental noise of each experiment. The latter is modeled as
\begin{equation}
    N_{\ell}^{XX} = \sigma_X^2\exp\left({\frac{\ell (\ell+1)\theta_{\textrm{\tiny{FWHM}}}^2}{8\ln 2}}\right)\;.
    \label{eq:12}
\end{equation}
Here, $\sigma_X$ represents the noise level for temperature and polarization, and $\theta_{\textrm{\tiny{FWHM}}}$ is the beam width, in units of radians. 

To produce and run around the fiducial model, we consider a CMB-S4 configuration, in which the sky fraction observed is $f_{\textrm{sky}}=0.4$. We take $\sigma_T=\sigma_{E}/\sqrt{2}=1\mu$K-arcmin, for perfect polarized detectors, and $\theta_{\textrm{\tiny{FWHM}}}=3$ arcmin to characterize the noise in Eq. (\ref{eq:12}), in agreement with \cite{Brinckmann:2018owf}. We also add the contribution of data from the LiteBIRD satellite, set to investigate the CMB polarization modes, with expected sky fraction coverage of $f_{\textrm{sky}}=0.7$. Considering information from the E and B spectra, we have noise characterized by $\sigma_T=\sigma_{E,B}/\sqrt{2}=4.17\mu$K-arcmin, and $\theta_{\textrm{\tiny{FWHM}}}=23$ arcmin, for a multipole range of $2<\ell< 200$. This will compose our baseline analysis, such that we can estimate the impact of CMB-only data on QI models.

\subsection{Euclid}
The Euclid galaxy survey has been collecting data since early 2024 and will cover a sky fraction of up to $ f_{\text{sky}} = 0.3636 $. Its first public data release is scheduled to take place in October 2026~\cite{euclidqr25}. For our analysis, we consider only the galaxy clustering component of the Euclid photometric (GC$_{\textrm ph}$) and spectroscopic (GC$_{\textrm sp}$) survey. 

\begin{itemize}
    \item Euclid photometric: For the GC$_{\textrm ph}$ case, the observable we use is the angular power spectrum $C^{GG}_{ij}(\ell)$, defined between redshift bins $i$ and $j$ as,
\begin{equation}
        C^{GG}_{ij}(\ell) = c \int_{z_{\min}}^{z_{\max}} \frac{dz}{H(z)r^2(z)} W^G_i(k_\ell, z) W^G_j(k_\ell, z) P^{GG}_{\delta \delta}(k_\ell, z) ~+~
    N^{GG}_{ij}(\ell),
\end{equation}
where $r(z)$ is the comoving distance, $H(z)$ is the Hubble parameter, and $P^{GG}_{\delta \delta}(k,z)$ is the nonlinear matter power spectrum at $k_\ell = (\ell + 1/2)/r(z)$. The window function $W^G_i(k,z)$ for the $i$-th redshift bin is given by,
\begin{equation}
    W^G_i(k, z) = \frac{H(z)}{c} b_i(k, z) \frac{n_i(z)}{\bar{n}_i},
\end{equation}
where $b_i(k, z)$ is the galaxy bias and $n_i(z)/\bar{n}_i$ is the normalized galaxy redshift distribution in the $i$-th bin. We assumed 10 redshift bins ranging from $z_{\min} = 0.001$ to $z_{\max} = 2.5$, with the total galaxy number density set to $n_{\text{gal}} = 30 \, \text{arcmin}^{-2}$. The average number of galaxies per bin per steradian is given by $\displaystyle \bar{n}_i = \frac{n_{\text{gal}}}{N_{\text{bins}}}$. The shot noise contribution to the auto-correlation is modeled as $\displaystyle N^{GG}_{ij}(\ell) = \frac{1}{\bar{n}_i} \delta_{ij}$.

For this analysis, we have adopted the maximum multipoles used in the likelihood analysis up to $\ell_{\max}^{\text{Gph}} = 750$, following the prescription described in~\cite{euclid2025_sensi}. The galaxy bias $b_i(k,z)$ is modeled with a nuisance parameter per redshift bin.

\item Euclid Spectroscopic: The Euclid spectroscopic survey provides three-dimensional clustering information of galaxies, modeled through the observed galaxy power spectrum,

\begin{equation}
P_{\text{obs}}(k_{\text{fid}}, \mu_{\text{fid}}; z) 
= \frac{1}{q_\perp^2(z) q_\parallel(z)}
\Bigg\{ \frac{\left[ b\sigma_8(z) + f(k, z)\sigma_8(z)\mu^2 \right]^2}{1 + f(k,z)^2 k^2 \mu^2 \sigma_p^2(z)} 
 \frac{P_{\text{dw}}(k, \mu; z)}{\sigma_8^2(z)} F_z(k,\mu;z) + P_s(z) \Bigg\}\nonumber \\
\quad
\end{equation}

where $f(k,z)$ is the linear growth rate, $P_{\text{dw}}(k,\mu;z)$ is the de-wiggled power spectrum, smoothed to account for BAO damping due to nonlinear effects, and $P_s(z)$ is the shot noise. Three main effects were considered in the modeling of the observed power spectrum: (I) the Alcock–Paczynski effect, due to the mismatch between the true and fiducial cosmology and is given by the first term on the observed power spectrum, $\frac{1}{q_\perp^2(z) q_\parallel(z)}$, where $q_\perp=\frac{D_A(z)}{D_{A,fid}(z)}$ and $q_\parallel=\frac{H_{fid}(z)}{H(z)}$; (II) the Kaiser effect, which accounts for redshift-space distortions and is given by $\left[b(z)\sigma_8(z) + f(k,z)\sigma_8(z)\mu^2\right]$, and (III) the Fingers-of-God effect, which arises from small-scale random motions within virialized structures, and appears on the denominator as $1 + f(k,z)^2 k^2 \mu^2 \sigma_p^2(z)$. The term $F_z(k,\mu;z)=\exp\left[-k^2\mu^2\sigma_r^2(z)\right]$ is the redshift error damping term, where $\sigma_r(z)=c\sigma_z/H(z)$ being $\sigma_z=0.002$ the redshift error.
\end{itemize}

For this work, we adopt the maximum scale up to $k_{\max} = 0.25 \, h/\text{Mpc}$, which follows the pessimistic spectroscopic configuration given by~\cite{euclid2025_sensi}, while using the likelihood available on MontePython, discussed in \cite{Euclid:2023pxu}.

\subsection{Model parameters}

As we are seeking to constrain the uncertainties on the $\alpha$ parameter, we choose to keep CMB as our baseline, so that it can be combined with Euclid projections. For simplicity, we only consider the impact of galaxy clustering, where we consider the spectroscopic and photometric likelihoods for Euclid. The CMB likelihood allows us to obtain important restrictions on the primordial parameters, namely $A_s$ and $n_s$, as well as on $\alpha$, given the unified nature of the model. Our fiducial model is then generated from the following cosmological parameters: $\omega_{cdm}=0.12$, $\omega_b=0.0237$, $h=0.675$, $\tau_{reio}=0.0544$, $n_s=0.965$, $\ln(10^{10}A_s)=3.0447$, and $\alpha=2$. This value of $\alpha$ is motivated by the recent analysis in \cite{Alestas:2024eic}, which constrains this parameter as $\alpha=1.89^{+0.40}_{-0.35}$, for a Planck+Pantheon Plus+DESI joint analysis. Another important parameter in quintessence models is the initial value of the scalar field, $\phi_{\textrm ini}$. As discussed in \cite{Akrami:2017cir,Akrami:2020zxw}, the values of $\phi_{\textrm ini}$ vary essentially due to the reheating mechanism that takes place after inflation. Particularly, for an exponential-type potential, it is found that the limit $\phi_{\textrm ini}=[-35,-10]$ covers the range between a gravitational reheating (lower limit) and more efficient mechanisms (upper limit). This is important, as it also impacts the late-time constraints of the model, since values of $\phi_{\textrm ini}$ closer to zero produce more significant deviations from the $\Lambda$CDM model, for a given $\alpha$. We then follow \cite{Akrami:2020zxw}, and fix the value of the initial field displacement, although we only consider here the $\phi_{\textrm ini}=-10$ case. On the other hand, we also include the impact of $\alpha$ as a free parameter. Finally, we also consider constraints on the derived parameters $M^2$, $\gamma$, and $r$. Their corresponding fiducial values are $M^2/M_P^4\times 10^{11}=3.985$, $\gamma=128.52,$ and $r=0.00735$.

\begin{table*}[t]
    \fontsize{7.5}{15}\selectfont
	\centering
	\begin{tabularx}{\textwidth}{ccccccccccc}
		\hline
        \hline
		\multicolumn{11}{c}{$\sigma$ (at 68\% CL)}\\
		\hline
        Experiment & $\omega_b$ & $\omega_{cdm}$ & $h$ & $\tau_{reio}$ & $n_s$ & $\ln 10^{10}A_s$ & $\alpha$ & $\gamma$ & $M^2/M_P^4\times 10^{11}$ & $r$ \\
        
        \hline

        CMB  & $0.000034$ & $0.00026$ & $0.0032$ & $0.002$ & $0.0015$ & $0.0034$ & $0.26$ & $0.195$ & $0.45$ & $0.00073$ \\        
        CMB+GC$_{\textrm ph}$ & $0.000032$ & $0.00023$ & $0.0030$ & $0.0019$ & $0.0013$ & $0.0033$ & $0.21$ & $0.16$ & $0.39$ & $0.00063$ \\
        CMB+GC$_{\textrm sp}$ & $0.000031$ & $0.00022$ & $0.0022$ & $0.0018$ & $0.0013$ & $0.0029$ & $0.17$ & $0.13$ & $0.34$ & $0.00055$  \\
        
        \hline 
		\hline
	\end{tabularx}
	\caption{1$\sigma$ uncertainty projections for the quintessential inflation model in Eq. \ref{eq:13}, for CMB (CMB-S4+LiteBIRD), CMB+GC$_{sp}$, CMB+GC$_{ph}$.}
	\label{tab:1}
\end{table*}

\section{Results and Conclusions}\label{sec4}

\begin{figure*}
    \centering
    \includegraphics[width=\columnwidth]{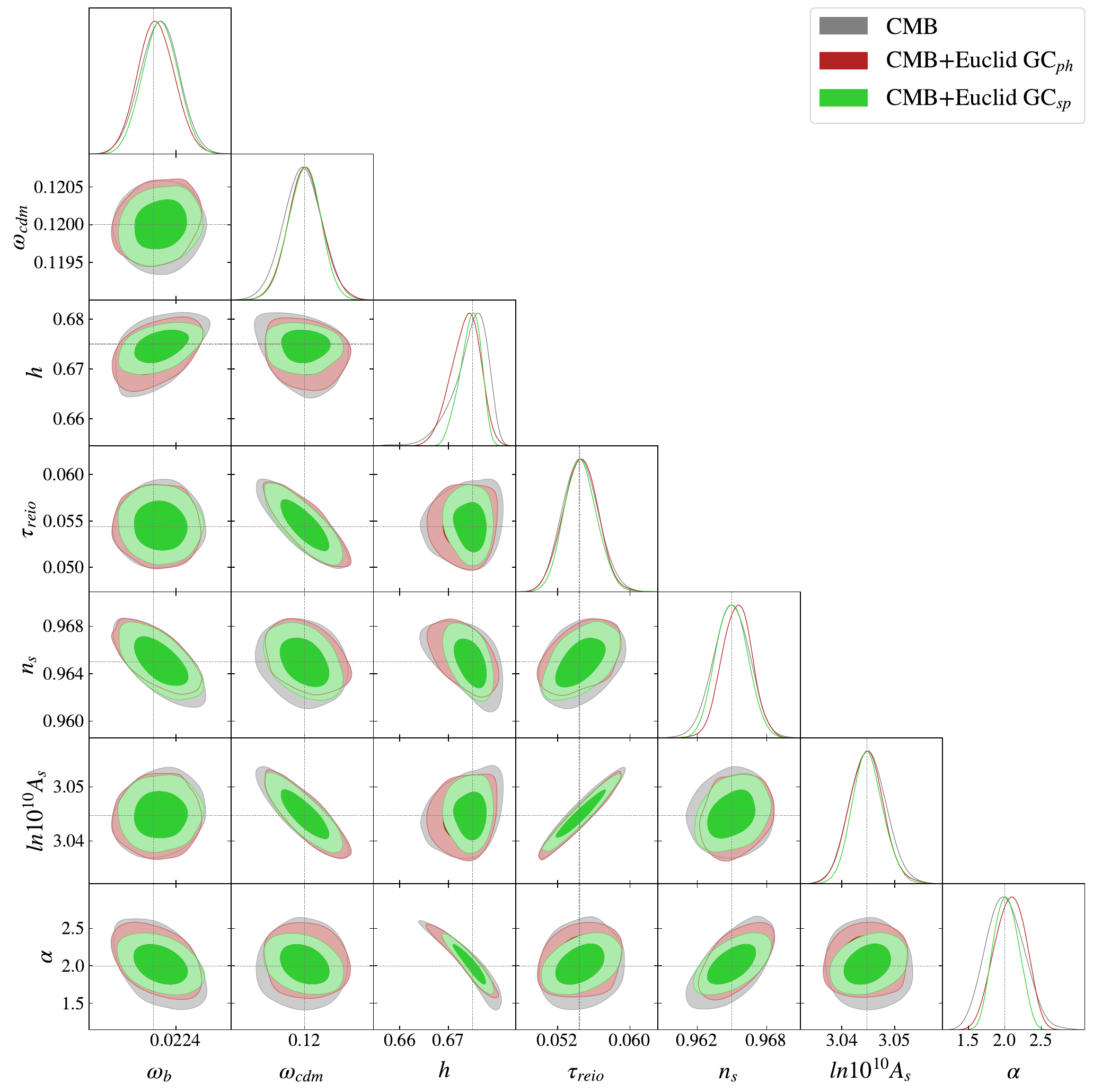}
    \caption{Posterior probability distribution at 68\% and 95\% confidence levels on the quintessential inflation model given by Eq. (\ref{eq:13}), as projected by CMB-S4+LiteBIRD (gray), CMB-S4+LiteBIRD+Euclid GC$_{\textrm sp}$ (green) and CMB-S4+LiteBIRD+Euclid GC$_{ph}$ (red).}
    \label{fig:1}
\end{figure*}

In this section, we present and discuss the results of the MCMC analysis for the sensitivities of the cosmological parameters under study, namely
\begin{equation}
    \{\omega_{cdm}, \omega_b, h, \tau_{reio}, n_s, A_s, \alpha\}.
\end{equation}
In Figure \ref{fig:1}, we present the $68.3\%$ and $95\%$ confidence regions for the model parameters, for the cases of CMB (Stage IV and LiteBIRD) only and the combination of the CMB with either GC$_{ph}$ or GC$_{sp}$. In Table \ref{tab:1}, we present the precision reached on all the aforementioned parameters when CMB-only probes are considered, and the combination of the former with Euclid.

As can be seen in the contour regions and Table \ref{tab:1}, using CMB only improves the constraints on all parameters by one order of magnitude compared with the results obtained in \cite{Alestas:2024eic}, which considered current CMB+lensing data. The most prominent improvements stand for $h$ and $\alpha$, where we have a gain of $103\%$ in the sensitivity for $h$, and constraints of $\mathcal{O}(10^{-1})$ for $\alpha$, which previously had only an upper limit of $<3$\footnote{The percentage of improvement is defined as $\displaystyle \left(\frac{\sigma_{\mathrm{before}}}{\sigma_{\mathrm{after}}}-1\right)\times 100\%$, where $\sigma_{\mathrm{before}}$ and $\sigma_{\mathrm{after}}$ in this case are, respectively, the 1$\sigma$ uncertainties found in \cite{Alestas:2024eic} and in this work by considering CMB only.}. In particular, this result for $\alpha$ was expected, since, as discussed in \cite{Akrami:2020zxw} and in the previous section, the combination of GC with CMB imposes simultaneous constraints on $n_s$ and $r$, allowing for precise measurements of the $\alpha$ parameter due to the relation $\displaystyle r=3\alpha(1-n_s)^2$.

Unsurprisingly, when combining CMB and Euclid, the reduction in the uncertainties on the parameters $h, \alpha, n_s, \omega_b, A_s$ are, respectively, $81.6\%$, $120.6\%$, $150\%$, $222.6\%$, and $469\%$, compared with the constraints found in \cite{Alestas:2024eic} by using current CMB, GC, and Supernovae Ia data\footnote{In this case, $\sigma_{\mathrm{before}}$ are the 1$\sigma$ uncertainties found in \cite{Alestas:2024eic} when considering the combination of CMB, GC and SNIa. And $\sigma_{\mathrm{after}}$ are the 1$\sigma$ uncertainties found in this work for the combination of CMB and GC$_{sp}$.}. The improvements mentioned above mean that the future surveys will yield sensitivities varying from $\mathcal{O}(10^{-5}-10^{-4})$ on the matter density and the tensor-to-scalar ratio parameters, to sensitivities of $\mathcal{O}(10^{-3}-10^{-1})$ for all the other parameters considered. This is remarkable since previous forecasts considering GC surveys~\cite{Akrami:2020zxw}, such as LSST~\cite{2019ApJ...873..111I} and SKA~\cite{{SKA:2018ckk}}, obtained the same order of sensitivity with fewer free parameters ($\alpha$ was kept fixed in those analyses). This level of precision will be enough to distinguish between the different classes of $\alpha$-attractor models indicated by the theoretically motivated values of $\alpha$, namely $\{1/3,2/3,1,4/3,5/3,2,7/3\}$.

\begin{figure*}
    \centering
    \includegraphics[width=0.47\columnwidth]{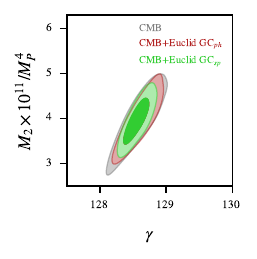}
    \includegraphics[width=0.5\columnwidth]{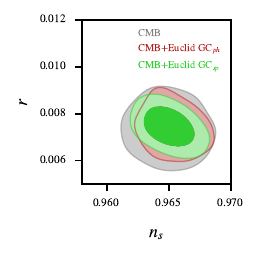}    
    \caption{Posterior probability distribution at 68\% and 95\% confidence levels on the parameters characterizing the $\alpha$-attractor potential Eq. \ref{eq:8} and the inflationary parameters $n_s$ and $r$ for the same model, as projected by CMB-S4+LiteBIRD (gray), CMB-S4+LiteBIRD+Euclid GC$_{\textrm sp}$ (green) and CMB-S4+LiteBIRD+Euclid GC$_{ph}$ (red). }
    \label{fig:2}
\end{figure*}

\begin{figure*}
    \centering
    \includegraphics[width=\columnwidth]{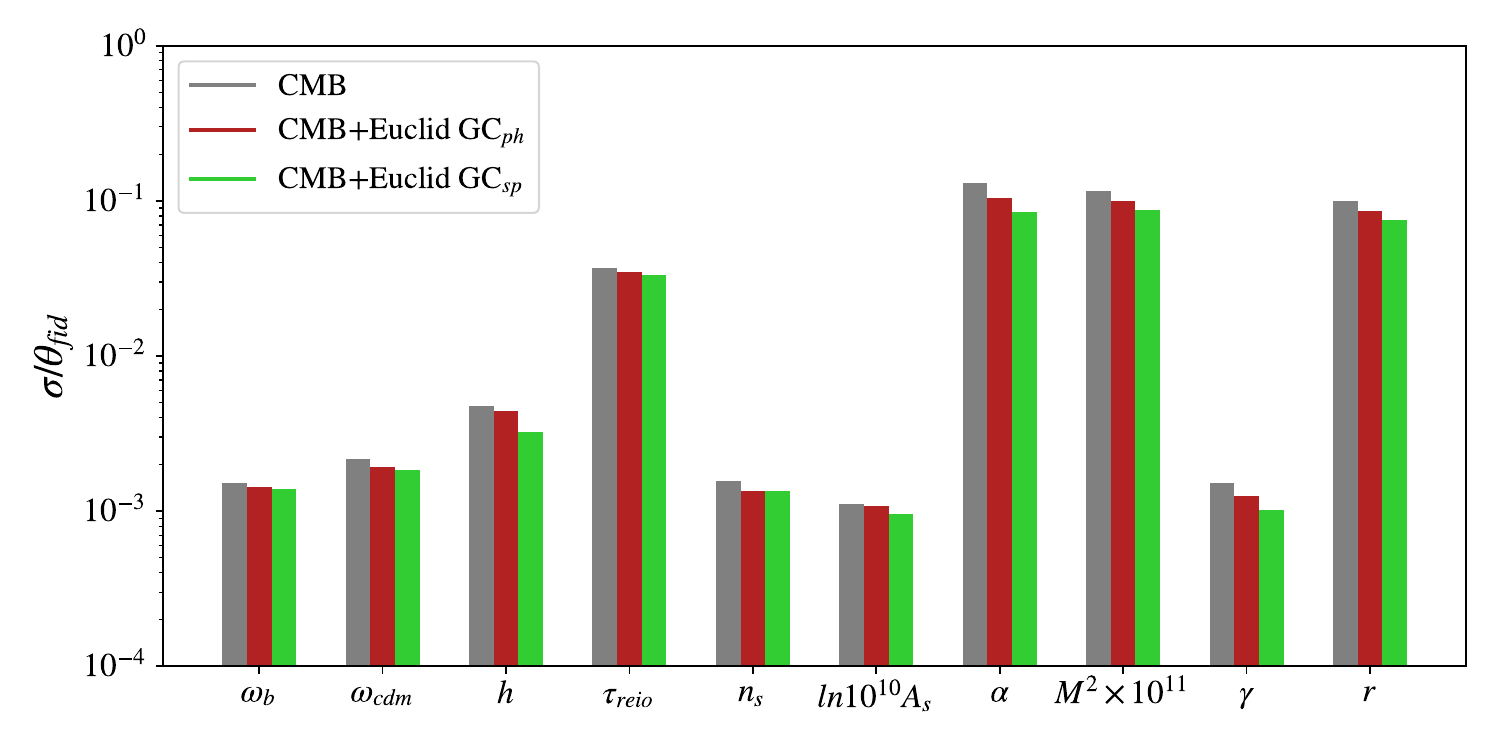}
    \caption{The ratio between the obtained uncertainties in Table \ref{tab:1} and fiducial values $\theta_{fid}$ of the parameters, for each experiment combination.}
    \label{fig:3}
\end{figure*}

Additionally, improvements are seen when combining the CMB with galaxy clustering data, whether from photometric or spectroscopic components of the Euclid survey. The uncertainties for the parameters \(n_s\), \(A_s\), \(r\), \(h\), and \(\alpha\) decrease by approximately \(7.1\%\), \(17.2\%\), \(32.7\%\), \(45.4\%\), and \(52.9\%\), respectively, when CMB is combined with GC$_{sp}$\footnote{In this case, $\sigma_{\mathrm{before}}$ are the 1$\sigma$ uncertainties found when considering the combination of CMB and GC$_{ph}$, and $\sigma_{\mathrm{after}}$ are the 1$\sigma$ uncertainties for the combination of CMB and GC$_{sp}$.}. 
In Figure \ref{fig:2}, we present the forecasted constraints for the other two parameters specific to the $\alpha$-attractor models: \(M^2\) and \(\gamma\). Notably, the most significant improvement is observed when combining the CMB with GC$_{sp}$, yielding reductions in the uncertainties of approximately \(15.4\%\) for \(M^2\) and \(21.9\%\) for \(\gamma\). 
These differences are visually represented in Figure \ref{fig:3}, which shows the ratio between the obtained uncertainties and the fiducial values of each parameter for different combinations of experiments.

The relationship between the parameters \(n_s\), \(H_0\), \(\alpha\), and \(\omega_{cdm}\) is of significant interest, particularly regarding their (anti-)correlation. Recent analyses of DESI BAO \cite{DESI:2024mwx} data suggest subtle alterations in matter clustering, which could lead to variations in the amplitude of scalar perturbations (\(A_s\)) and spectral index (\(n_s\))\cite{daCosta:2024grm}. Notably, in our investigation, we observed a marked improvement in the constraints for \(\omega_{cdm}\), achieving a reduction by more than three orders of magnitude compared to existing constraints \cite{Alestas:2024eic}. This finding holds substantial importance since previous studies, such as \cite{Pedrotti:2024kpn}, indicate that if both the matter density parameter \(\Omega_m\) and the baryon density parameter \(\omega_b\) are calibrated — either through BAO measurements or uncalibrated supernovae (SNeIa) and Big Bang Nucleosynthesis (BBN) — then an increase in the Hubble constant (\(H_0\)) necessitates a corresponding rise in \(\omega_{cdm}\). Therefore, the anticipated enhancements in the sensitivity of \(\omega_{cdm}\) from future galaxy cluster surveys are expected to provide deeper insights into the Hubble tension, possibly illuminating the underlying mechanisms associated with it. Additionally, the projected constraints on \(n_s\) and the tensor-to-scalar ratio (\(r\)) will be robust enough to determine whether a quintessential or non-quintessential model is required to account for cosmic inflation.

In summary, our findings highlight the significant impact that upcoming large-scale structure and CMB surveys will have in refining the parameters of the $\alpha$-attractor quintessential inflation. These surveys are expected to achieve a higher level of sensitivity, allowing for much stricter constraints on key parameters associated with $\alpha$-attractor models, including the scalar spectral index $n_s$ and the curvature of the K\"{a}hler manifold associated with $\alpha$. The former,  particularly, provides a robust method to differentiate between quintessential $\alpha$-attractor models and their non-quintessential counterparts, once the existence of a kination phase on the quintessential inflation scenarios changes the prediction for this parameter in relation to non-quintessential ones, as pointed out in \cite{Akrami:2020zxw}. Moreover, the anticipated strong constraints on the matter density parameter will enhance our understanding of existing tensions within the $\Lambda$CDM framework, particularly by exploring the relationship between this parameter and the Hubble constant.

\section*{Acknowledgements}

We acknowledge the use of high-performance computing services provided by the Observatório Nacional Data Center. GR is supported by the Coordena\c{c}\~ao de Aperfei\c{c}oamento de Pessoal de N\'ivel Superior (CAPES). FBMS is supported by Conselho Nacional de Desenvolvimento Científico e Tecnológico (CNPq) grant No. 151554/2024-2. S.S.C. acknowledges support from the Istituto Nazionale di Fisica Nucleare (INFN) through the Commissione Scientifica Nazionale 4 (CSN4) Iniziativa Specifica ``Quantum Fields in Gravity, Cosmology and Black Holes'' (FLAG) and from the Fondazione Cassa di Risparmio di Trento e Rovereto (CARITRO Foundation) through a Caritro Fellowship (project ``Inflation and dark sector physics in light of next-generation cosmological surveys''). JGR acknowledges financial support from the Funda\c{c}\~ao de Amparo \`a Pesquisa do Estado do Rio de Janeiro (FAPERJ) grant No. E-26/200.513/2025. RvM is suported by Fundação de Amparo à Pesquisa do Estado da Bahia (FAPESB) grant TO APP0039/2023. RS  acknowledges financial support from CNPq (Grant no. 307630/2023-4). DFM thanks the Research Council of Norway for their support and the resources provided by UNINETT Sigma2 -- the National Infrastructure for High-Performance Computing and Data Storage in Norway. JSA is supported by CNPq No. 307683/2022-2 and FAPERJ grant 259610 (2021).

\bibliography{references}

\end{document}